\documentstyle[pra,aps,psfig,preprint,tighten]{revtex}

\begin{document}

\draft

\preprint{Revision 5: 4/2/97}

\title{Mean-field analysis and Monte Carlo study of an
interacting two-species catalytic
surface reaction model}

\author{K. S. Brown, K. E. Bassler, D. A. Browne}

\address{
Department of Physics and Astronomy,
Louisiana State University,
Baton Rouge, Louisiana 70803}

\date{\today}

\maketitle
\begin{abstract}

We study the phase diagram and critical behavior of an interacting one
dimensional two species monomer-monomer catalytic surface reaction
model with a reactive phase as well as two equivalent adsorbing phase
where one of the species saturates the system.  A mean field analysis
including correlations up to triplets of sites fails to reproduce the
phase diagram found by Monte Carlo simulations.  The three phases
coexist at a bicritical point whose critical behavior is described by
the even branching annihilating random walk universality class.  This
work confirms the hypothesis that the conservation modulo 2 of the
domain walls under the dynamics at the bicritical point is the
essential feature in producing critical behavior different from
directed percolation.  The interfacial fluctuations show the same
universal behavior seen at the bicritical point in a three-species
model, supporting the conjecture that these fluctuations are a new
universal characteristic of the model.

\end{abstract}


\narrowtext

\section{Introduction}

Nonequilibrium statistical models with many degrees of freedom whose
dynamics violate detailed balance arise in many areas such as
biological populations, chemical reactions, fluid turbulence, traffic
flow, and growth/deposition processes.  The macroscopic behavior of
these models can be much richer than that of systems in thermal
equilibrium, showing organized macroscopic spatial and temporal
structures like pulses or waves, and even spatiotemporal chaos.  Even
the steady state behavior of a homogeneous system without these
structures can be far more complicated, involving for example a scale
invariant steady state without tuning the system to a specific point.
However, like their equilibrium cousins, systems at continuous
transitions between nonequilibrium steady states show universal
behavior that is insensitive to microscopic details and depends only on
properties such as symmetries and conservation laws.

One class of models that have received extensive study are those with
absorbing phase transitions where the system changes from an active
state with statistical fluctuations about the mean behavior to a
noiseless inert state consisting of a single microscopic
configuration.  The term absorbing refers to the fact that the system
cannot leave this state once it reaches it.  Examples include directed
percolation (DP)\cite{DP1,DP2}, the contact process\cite{CP},
auto-catalytic reaction models\cite{ABW}, and branching annihilating
random walks with odd numbers of offspring\cite{BAW2,BAW3}.  Both
renormalization group calculations\cite{DP1,RFT} and Monte Carlo
simulations\cite{DP2,CP,ABW,BAW2,BAW3,ZGB2} show that these models form
a single universality class for a purely nonequilibrium model with no
internal symmetry in the order parameter.

Recently, a number of models with continuous adsorbing transitions in a
universality class distinct from directed percolation have been
studied.  These models include probabilistic cellular automata models
studied by Grassberger and coworkers\cite{PCA}, certain kinetic Ising
models\cite{NKI}, the interacting monomer-dimer model\cite{IMD}, a
three species monomer model with Potts-like
symmetry\cite{shortBB,longBB} and branching annihilating random walks
with an even number of offspring (BAWe)\cite{BAW2,BAWe}.  All of these
models except for the BAWe have two equivalent absorbing states
indicating the importance of symmetry of the adsorbing state to the
universality class.  However, the universal behavior of this new class
is apparently controlled by a dynamical conservation law.  If the
important dynamical variables in this class are defects represented by
the walkers in the BAWe model and the walls between different saturated
domains in the other models, the models have a ``defect parity''
conservation law\cite{PCA} where the number of defects is conserved
modulo 2.  Recent field theoretic work confirms this
viewpoint\cite{CardyRG}.

Recently, two of us\cite{shortBB,longBB} investigated a three-species
monomer model with annihilation reactions between dissimilar species.
The transitions from the reactive state to a single absorbing state
where one species saturates the system fell in the DP universality
class.  These phase boundaries meet at bicritical points\cite{bicrit}
where two different absorbing states coexist.  Because the domain walls
between different domains of the two phases spawn and annihilate in
pairs, the critical behavior fell into the BAWe universality class.  We
also showed that at the bicritical point the characteristic
fluctuations of the domain walls between the equivalent absorbing
phases was given by a new exponent.

The present paper studies the connection between the behavior at a
bicritical point and the presence of the BAWe critical behavior and
tests the universality of the interfacial fluctuations.  We investigate
a model introduced by Zhuo, Redner, and Park\cite{Redner1} that also
has a bicritical point at the junction of two absorbing phase
transitions.  We find that the bicritical exponents in this model also
fall into the BAWe class and the interfacial exponents are the same as
in the three species model, lending credence to the universal nature
of the interfacial fluctuations.

In the next section we introduce the model and present results for the
mean field theory in two levels of approximation.  Section 3 discusses
our static and dynamic Monte Carlo simulations that yield a different
phase diagram from the mean field results and produce critical
exponents for the dynamic behavior that fall into the BAWe class.  The
final section contains our conclusions.

\section{The Model}

\begin{figure}[p]
\centerline{\psfig{file=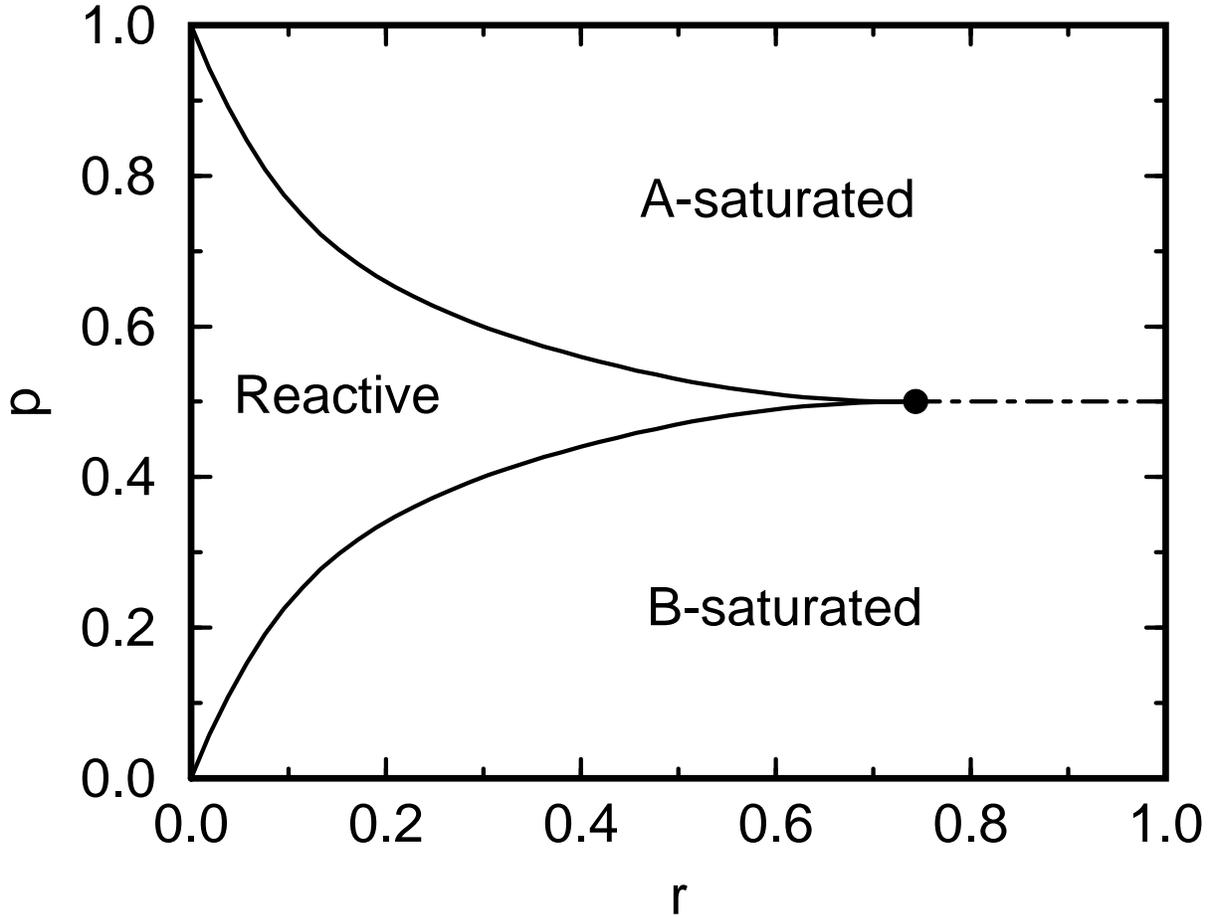,width=6.4truein}}
\caption{Phase diagram determined by Monte Carlo simulation for the
interacting monomer-monomer model.  The continuous transitions from
either of the saturated phases (A and B) to the reactive phase are
shown as solid lines, while the transition between the two saturated phases
(dot-dash line) is first-order.  The three phases coexist at a
bicritical point at $(r_c,p_c)\!=\!(0.7435,0.5)$, which is shown as a
heavy dot.}
\label{phasediag}
\end{figure}

The model we study was first introduced by Zhuo, Park and
Redner\cite{Redner1}.  Two monomers, called $A$ and $B$, adsorb at the
vacant sites of a one-dimensional lattice with probabilities $p$ and
$q$, respectively, where $p+q=1$.  The adsorption of a monomer at a
vacant site is affected by monomers present on neighboring sites.  If
either neighboring site is occupied by the same species as that trying
to adsorb, the adsorption probability is reduced by a factor $r<1$,
mimicking the effect of a nearest-neighbor repulsive interaction.
Unlike monomers on adjacent sites react immediately and leave the
lattice, leading to a process limited only by adsorption.  We have
performed static Monte Carlo simulations that produce a phase diagram
similar to the one found by Zhuo {\em et al.}\cite{Redner1}.  The
diagram, displayed in Fig.~\ref{phasediag} with $p$ plotted {\em vs.\/}
$r$, shows a reactive steady state (R) bordered by two equivalent
saturated phases (labeled A and B).  The transitions from the reactive
phase to either of the saturated phases are continuous, while the
transition between the saturated phases is first-order discontinuous.
The two saturated phases meet the reactive phase at a {\em
bicritical\/} point\cite{bicrit} at a critical value of $r=r_c$.  In
the case of $r = 1$, the reactive region no longer exists\cite{MM,MMar}
and the only transition line is the first-order discontinuous line
between the saturated phases.  Another model in which the adsorption
repulsion is not symmetric (only A's feel the repulsion) has been
studied \cite{Redner1}, and its critical behavior was found to be in
the DP universality class.  However, no effort was made to determine
either the location of the bicritical point or the bicritical behavior
of the model.

\section{Mean-field theory}

To analyze the kinetics of this model, it is useful to perform a
mean-field analysis.  While such analysis neglects long-range
correlations and thus cannot be expected to properly predict critical
properties, it should properly predict the qualitative structure of the
phase diagram, including the existence of continuous transitions and
multi-critical points.  The mean-field analysis also provides a
starting point for studying the importance of such fluctuations, which
become particularly important near continuous phase transitions.  The
mean-field approach we use\cite{DickmanMFT} studies the time evolution
of clusters of sites, the approximation coming in truncating the
probabilities of observing clusters of larger size into probabilities
for smaller size clusters.  The analysis presented below of this 1D
model includes clusters consisting of up to triplets of adjacent
sites.

At a particular time, a lattice with $N$ sites will have $N_V$
vacancies, the remaining sites being filled with $N_A$ A monomers and $N_B$
$B$ monomers.
The density of species $i = A$, $B$, $V$ is $x_i \equiv (N_i/N)$.
We have the obvious constraint
\begin{equation}
x_V + x_A + x_B = 1,
\label{siteconstraint}
\end{equation}
so we can take $x_A$ and $x_B$ as independent.

We next consider clusters of pairs of sites, where we define $N_{ij}$
as the number of pairs with species $i$ on one site and $j$ on the site
to its right, and the pair density $x_{ij}\equiv (N_{ij}/N)$.  Because
of the immediate reaction of $AB$ pairs we have $N_{AB}=N_{BA}=0$.
Using relations between the pair and site occupancies
\begin{eqnarray}
x_{A} & = & x_{AA}+x_{AV}=x_{VA}+x_{AA}\nonumber\\
x_{B} & = & x_{BB}+x_{BV}=x_{VB}+x_{BB}\nonumber\\
x_{V} & = & x_{AV}+x_{BV}+x_{VV},
\label{pairconstraint}
\end{eqnarray}
only two of the pair densities are independent of the $x_i$, which we
choose to be $x_{AA}$ and $x_{BB}$.

For the clusters of triples we define $N_{ijk}$ as the number of triples
with species $j$ in the central site, species $i$ to the left, and species
$k$ to the right, with the corresponding densities $x_{ijk}=(N_{ijk}/N)$.
The prohibition of adjacent $AB$ pairs and the relations between
the numbers of clusters of triples and pairs of sites

\begin{equation}
\begin{array}{rclcl}
x_{AA} & = & x_{AAV} + x_{AAA}           & = & x_{VAA} + x_{VVV}\\
x_{AV} & = & x_{AVV} + x_{AVA} + x_{AVB} & = & x_{VAV} + x_{AAV}\\
x_{VA} & = & x_{VVA} + x_{AVA} + x_{BVA} & = & x_{VAV} + x_{VAA}\\
x_{BB} & = & x_{BBV} + x_{BBB}           & = & x_{VBB} + x_{BBB}\\
x_{BV} & = & x_{BVV} + x_{BVA} + x_{BVB} & = & x_{VBV} + x_{BBV}\\
x_{VB} & = & x_{VVB} + x_{AVB} + x_{BVB} & = & x_{VBV} + x_{VBB}\\
x_{VV} & = & x_{VVV} + x_{VVA} + x_{VVB} 
       & = & x_{VVV} + x_{AVV} + x_{BVV} 
\end{array}
\label{tripleconstraint}
\end{equation}

gives a total of 6 independent triple densities: $x_{AAA}$, $x_{AVA}$,
$x_{BBB}$, $x_{BVB}$, $x_{AVB}$, and $x_{BVA}$.  The last two are equal
in a homogeneous steady state.

We must calculate the rate that each of these densities change due to
four allowed kinetic processes: deposition of an $A$ without reaction,
removal of an $A$ by reaction with a $B$ adsorbing next to it,
deposition of a $B$ without reaction, and removal of a $B$ by reaction
by an adsorbing $A$.  These four processes are detailed in
Table~\ref{ratetable}.  Because of the symmetry of the reaction rules,
we will usually only present explicit formulae for half of the
equations; the remainder can be found by interchanging $A$ and $B$
everywhere in the equations.

The exact equations for the single site densities are
\begin{eqnarray}
{dx_{A}\over dt}
& = & p\, x_{VVV} + p\,r\, (x_{VVA}+x_{AVV}+x_{VAV})\nonumber\\
&   & \mbox{} -q\, (x_{VVA}+x_{AVV}+x_{VAV})\nonumber\\
&   & \mbox{} -q\,r\,(x_{AVB}+x_{BVA})\nonumber\\
{dx_{B}\over dt}
& = & q\, x_{VVV} + q\,r\, (x_{VVB}+x_{BVV}+x_{VBV})\nonumber\\
&   & \mbox{} -p\, (x_{VVB}+x_{BVV}+x_{VBV})\nonumber\\
&   & \mbox{} -p\, r\,(x_{BVA}+x_{AVB})
\label{siteode}
\end{eqnarray}
which clearly shows the $A\leftrightarrow B$ symmetry noted above.

By considering each of the possible reactions in Table~\ref{ratetable},
the equation for the $AA$ pairs is
\begin{eqnarray}
{dx_{AA}\over dt}
& = & p\,r\,(x_{VVA} + x_{AVV} + 2x_{AVA})\nonumber\\
&   & \mbox{} -q\,(x_{VVAA} + x_{AAVV} +
            {1\over2} x_{AAVA} + {1\over 2} x_{AVAA})\nonumber\\
&   & \mbox{} -q\, r\,(x_{AAVB}+x_{BVAA})
\label{pairode}
\end{eqnarray}
where we have denoted densities of quartets of sites in an obvious
extension of our notation.

Finally the triples obey the equations
\widetext

\begin{eqnarray}
{dx_{AAA}\over dt}
& = & p\,r\,(x_{VVAA} + x_{AAVV} + x_{AVA} + x_{AVAA} + x_{AAVA})\nonumber\\
&   & \mbox{} -q\,(x_{VVAAA} + x_{AAAVV} +
            {1\over2} x_{AAAVA} + {1\over 2} x_{AVAAA})\nonumber\\
&   & \mbox{} -q\,r\,(x_{AAAVB}+x_{BVAAA})\nonumber\\
{dx_{AVA}\over dt}
& = & p\,(x_{VVVA} + x_{AVVV}) -p\,r\,x_{AVA}\nonumber\\
&   & \mbox{} -q\,(x_{VVAVA} + x_{AVAVV} + x_{AVAVA} + x_{AVA})\nonumber\\
&   & \mbox{} -q\,r\,(x_{AVAVB}+x_{BVAVA})\nonumber\\
{dx_{AVB}\over dt}
& = & p\,x_{VVVB} + q\,x_{AVVV} \nonumber\\
&   & \mbox{} -q\,(x_{VVAVB} + {1\over2}x_{AVAVB})
      -p\,(x_{AVBVV} + {1\over2}x_{AVBVB}) \nonumber\\
&   & \mbox{} -q\,r\,(x_{AVB} + x_{BVAVB}) -p\,r\,(x_{AVB} + x_{AVBVA})
\label{tripleode}
\end{eqnarray}
\narrowtext
In general, for a cluster of $M$ sites, the adsorption processes link
clusters of $M$ sites to clusters of $M\!+\!1$ sites and the reaction
processes link to clusters of $M\!+\!2$ sites.

\begin{figure}[p]
\centerline{\psfig{file=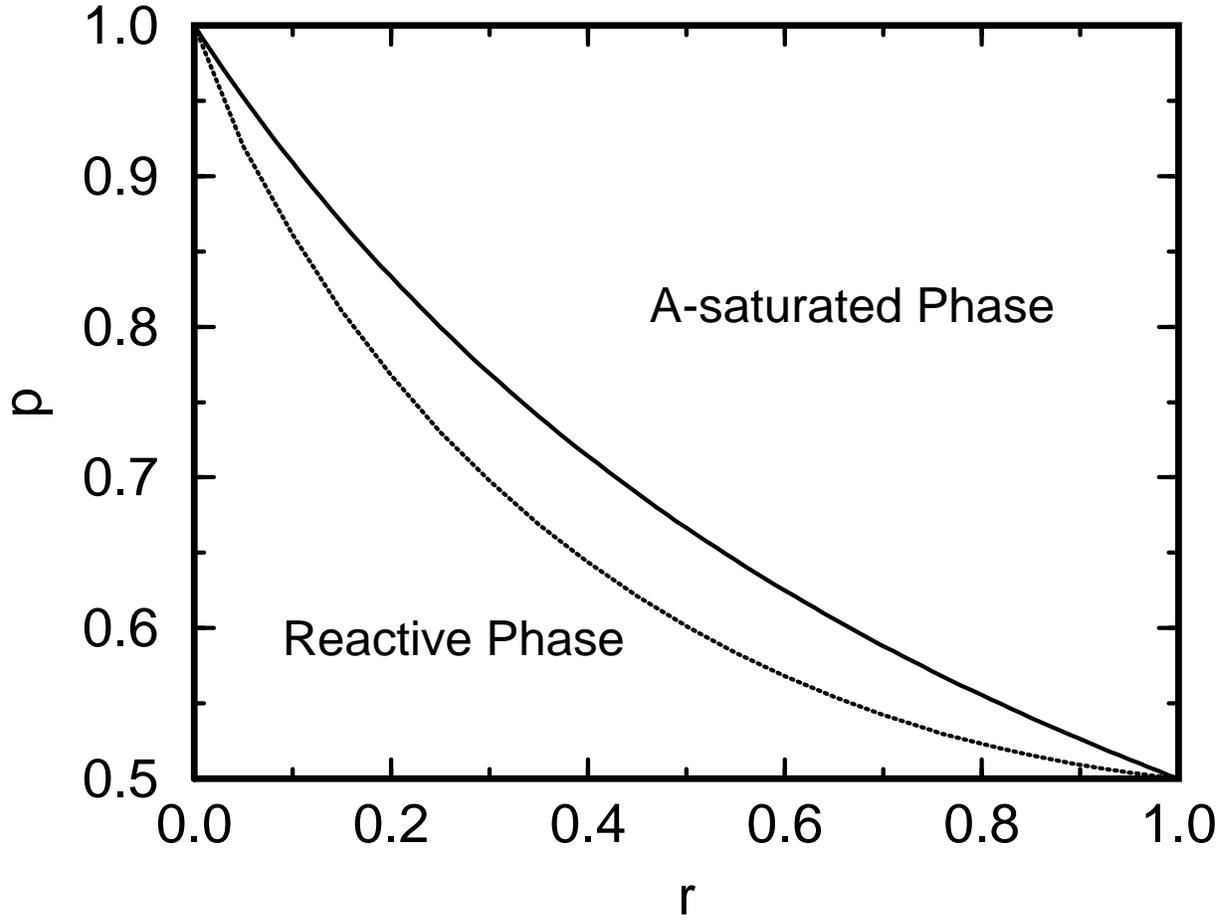,width=6.4in}}
\caption{Phase boundary between the A-saturated and reactive phases in
the site approximation (solid line) and pair approximation (dashed
line).  The triple approximation is indistinguishable from the pair
approximation on this figure.}
\label{meanfld}
\end{figure}

To close the equations we perform a truncation.  In the site
approximation, we ignore any spatial correlations and solve just
Eqn.~(\ref{siteode}) --- and its partner for $x_B$ -- using
Eqn.~(\ref{siteconstraint}) and the approximation $x_{ijk}\approx
x_{i}x_{j}x_{k}$.  We find the boundary between the saturated $A$ phase
and the reactive phase by finding the combination of parameters $p$ and
$r$ where the saturated phase fixed point loses its stability.  This
occurs when $p=1/(1+r)$.  The $B$-saturated phase boundary is given
by $q=1-p=1/(1+r)$.  These are shown as dashed lines in
Fig.~\ref{meanfld}.  The absorbing state transitions are continuous
but there is no coexistence line between the saturated phases, just
as we found in Ref.~\onlinecite{longBB}.

In the pair approximation, we replace correlations of triples and
larger clusters of sites by approximating them as
$x_{ijk}=x_{ij}(x_{jk}/x_{j})$,
and $x_{ijkl}=x_{ij}(x_{jk}/x_j)(x_{kl}/x_k)$.  Each term in
parentheses like $(x_{jk}/x_j)$ has the simple interpretation as the
conditional probability of finding species $k$ given that the site to
the left is occupied by species $j$.  We again find the phase diagram
by determining where the stability of the poisoned phase fixed point of
the system of Eqns.~(\ref{siteode}) and (\ref{pairode}) vanishes.
Unlike the site approximation, solving the equations to determine
the phase boundary was determined
numerically.  This procedure required some care because, although terms like $x_{VA}$ and $x_{V}$ vanish on
the phase boundary, their ratio, which appears in Eqns.~(\ref{siteode})
and (\ref{pairode}) in our approximation, is nonzero.  We note that,
just as for the three species model\cite{longBB}, Fig.~\ref{meanfld}
shows that this approximation fails to produce the coexistence curve
between the saturated phases.  This indicates that the large domains of
$A$ and $B$ that appear close to the absorbing phase transitions are
not accurately represented in the site and pair approximations.  In both
of these approximations, the width of the reactive window is proportional
to $(1-r)$ as $r$ approaches one.

The triple approximation is similar in spirit to the pair approximation
with higher order correlations being decomposed as conditional
probabilities via $x_{ijkl}\approx x_{ijk}(x_{jkl}/x_{jk})$ and
$x_{ijklm}\approx x_{ijk}(x_{jkl}/x_{jk})(x_{klm}/x_{kl})$.  The
numerically determined phase diagram is shown in Fig.~\ref{meanfld} 
and shows little difference from the pair approximation except
very close to $r=1$.  Surprisingly, the triple approximation fails
here to move the bicritical point away from $r=1$, which did happen
in our earlier work\cite{longBB} on the three-species model.  Very close to
$r=1$, the width of the reactive region varies as $(1-r)^{3/2}$
in this approximation.

We stop the mean field approach at the triple approximation because the
increase in algebra is not compensated by an improvement of the phase
diagram.  The failure of this mean field approach to produce a
realistic position for the bicritical point is linked to the fact that
the probability of observing a long cluster of sites all filled with
one species, which is approximated here by the probability of a smaller
cluster raised to a power, will decay exponentially with the size of
the cluster.  However, the simulations presented in this paper and
previous work\cite{shortBB,longBB} clearly show that at the bicritical
point large domains of each species are present in the steady state,
with the fundamental dynamical variables being the domain walls.

\section{Simulations}

The behavior of this model at the critical lines has been studied and
found to be in the DP universality class\cite{Redner1}, which we
verified using $10^5$ independent runs of up to $10^4$ time steps.
What has not been studied is the behavior  at the bicritical point.  We
use two types of ``epidemic'' simulations\cite{BAW3,BAWe,DPepi} to
study the system's behavior at the bicritical point.  The first (defect
dynamics)  starts with the lattice in a configuration close to one of
the saturated phases, in which a lone vacancy is surrounded by members
of one species.  The second (interface dynamics) studies the time
evolution of a minimum-width interface between the two saturated phases
(a lone vacancy is bordered on one side by A's and by B's on the other
side).  In the defect dynamics, the simulations are over when the
lattice becomes totally saturated.  For the interface dynamics the end
of the simulation is marked by the ``collapse'' of the interface back
to its minimum width\cite{shortBB}.

Since the total number of possible adsorption sites is usually very
small, we use a variable time algorithm to improve computational
efficiency.  A list of possible adsorption sites is kept during the
simulation, and one of these sites is picked at random for adsorption
according to $\lbrace p_\alpha \rbrace$ and, if applicable, $r$.  Time
is then incremented by $1/n_V(t)$, where $n_V(t)$ is the number of
vacancies in the lattice at that time.  To avoid end effects, we always
start with a lattice that is so large as to be considered of infinite
extent.

During the defect dynamics simulations, we measure $P(t)$, the
probability that the system will not become saturated at time $t$,
$n_V(t)$, the number of vacancies in the lattice at time $t$, $n_o(t)$,
the number of species opposite to those of the initial saturated
configuration at time $t$, and $R^2(t)$, the square size of the
reactive region in a surviving run at time $t$.  At a continuous phase
transition as $t \to \infty$ these quantities obey power laws
\begin{equation}
\begin{array}{cccc}
P(t) \propto t^{-\delta}& \langle R^2(t) \rangle \propto t^z&
\langle n_V(t) \rangle \propto t^{\eta} &
\langle n_o(t) \rangle \propto t^{\eta_o}
\end{array}
\end{equation}

The exponent $\eta_o$ is not independent of the others, but gives a
useful check on the calculation.  The scaling law for $\eta_o$ can be
understood by considering the set of all surviving runs.  The width of
the defect region grows as $t^{z/2}$.  This region is filled with
vacants and sites occupied by either of the species, so $[R^2(t)]^{1/2}
= c(n_V+n_A+n_B)P(t)$, with $c$ some constant.   The factor $P(t)$
accounts for the fact only the surviving runs contribute to $R^2(t)$.
If we further assume that in this defect (which is very large for long
times since $z>0$) we have the same kind of configurations as we would
see in a static simulation, we should have $n_o = N_A(t)=N_B(t)$.  Thus
we should see
\begin{equation}
2n_o(t) = {[R^2(t)]^{1/2}\over cP(t)} - n_V(t)
\sim c_1t^{z/2-\delta} - c_2t^{\eta}
\label{etaolaw}
\end{equation}
so we see that $\eta_o=(z/2)-\delta$.

For the interface dynamics simulations, we define $P(t)$ to be the
probability of the interface of avoiding collapse back to minimum
width, with $n(t)$ the vacancy concentration and $R^2(t)$ the
square-size of the interface, both similar to the above.  $n_o(t)$ has
no meaning or corresponding equivalent in this type of simulation, so
it is not measured.  As $t \to \infty$ these quantities obey power laws
\begin{equation}
P(t) \propto t^{-\delta'},\quad \langle n(t) \rangle \propto t^{\eta'},
\quad \langle R^2(t) \rangle \propto t^{z'}.
\end{equation}

Log-log plots of these measured quantities as a function of time are
straight lines at a phase transition and show curvature away from
transitions.  A precise estimate for both the location of the
bicritical point and the values of the exponents can be obtained by
examining the local slopes of the curves on a log-log plot.  For example,
the effective exponent $-\delta(t)$ is defined as
\begin{equation}
-\delta(t) = \lbrace \ln \lbrack P(t)/P(t/b) \rbrack / \ln b \rbrace,
\end{equation}
with similar definitions for $\eta(t)$, $\eta_o(t)$, and $z(t)$.  We
usually choose $b=5$ for our numerical work, with other values like
$b=2$ and $b=3$ being used to check the results.  At the bicritical
point, the value of the effective exponent should extrapolate to the
bicritical value in the long time limit ($t^{-1}\to0$).  Away from the
bicritical point, the local slope will show strong upward or downward
curvature as $t^{-1} \to 0$.

To determine the value of both the critical exponents and their
uncertainties, we use a technique different from that typically
employed\cite{BAW3,DPscaling}.  The data collected at the bicritical
point is divided into ten data sets that are statistically
independent.  A linear regression is then performed on each set, with
the intercept the only pertinent quantity.  The value for the exponent
is then calculated as the simple mean of the ten intercepts, and the
uncertainty is the standard error of the mean.  All of the results
quoted in this paper were obtained using this technique, which gives a
more unbiased estimate of the uncertainty than previous methods which
used the quality of the linear fit to the entire data set.

\begin{figure}[p]
\centerline{\psfig{file=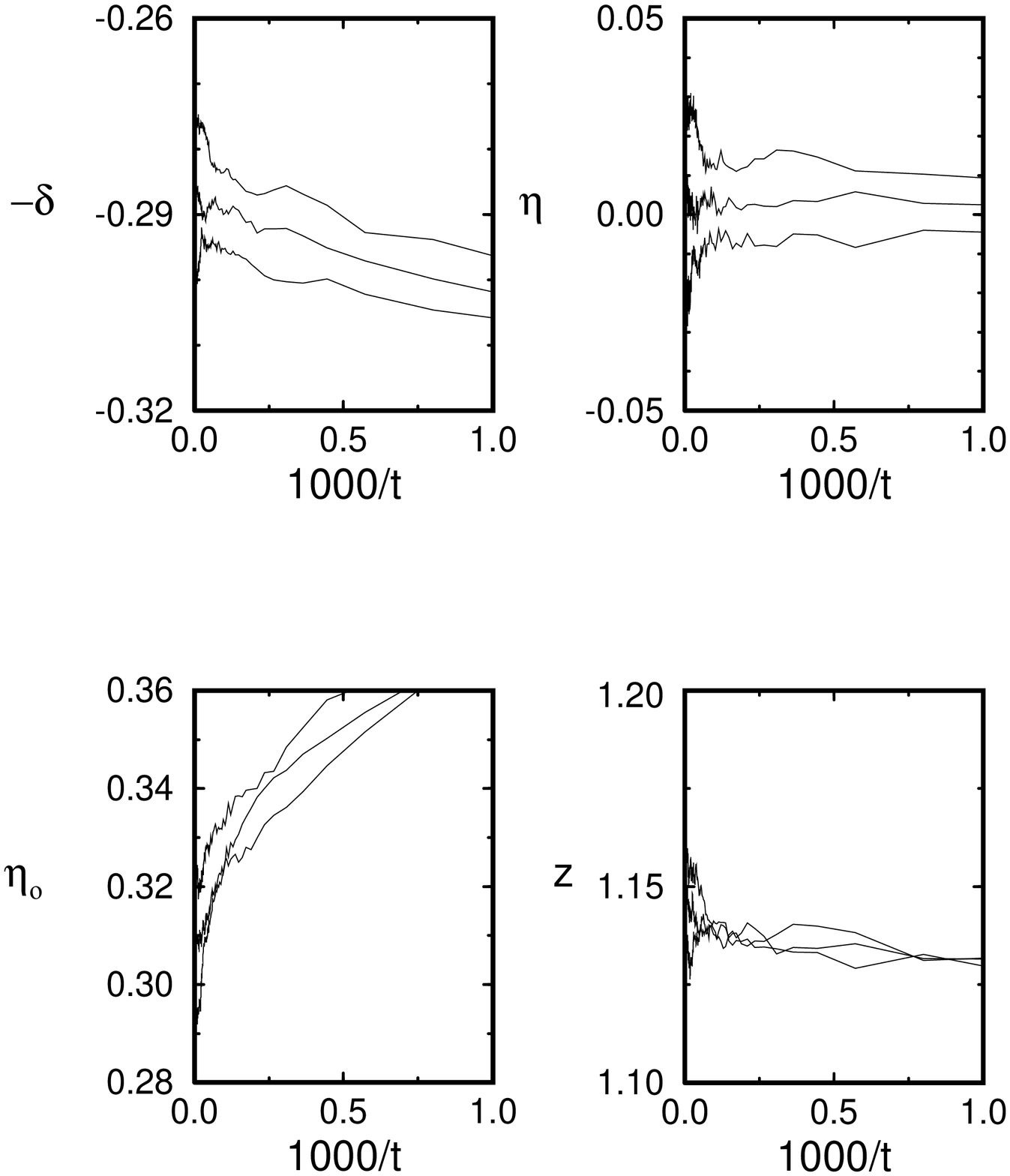,width=6.4truein}}
\caption{Effective exponents for the bicritical defect dynamics.  The three
curves correspond to three values of $r$ near the bicritical point.  From
top to bottom they are $r = 0.7420,0.7435,0.7450$, with the middle curves
corresponding to the bicritical values.}
\label{defect}
\end{figure}

Figure~\ref{defect} shows the local slopes of the data for the defect
dynamics simulations for $p = 0.5$ and $r$ near the bicritical
value.  The system started near an A saturated phase, but the symmetry
of the model dictates that we could equally as well have started near a
B saturated phase.  This data was calculated from $10^6$ independent
runs of up to $10^5$ time steps at each $r$ value.

The analysis presented in Fig.~\ref{defect} gives a critical repulsion
value of $r_c = 0.7435(15)$.  We find values of $\delta = 0.287(4)$,
$\eta = 0.003(4)$, $\eta_o = 0.306(5)$, and $z = 1.146(8)$ for the
critical exponents.  From these values we determine the bicritical
behavior falls in the BAWe universality class, for which\cite{BAWe} the
exponents are $\delta = 0.285(2)$, $\eta = 0.000(1)$, and $z =
1.141(2)$.  The value of $\eta_o$ does not agree very well with the
scaling law $\eta_o = (z/2)-\delta$ derived above, but that is the
result of the effect of the second term in Eq.~(\ref{etaolaw}), which
acts as a large correction to the asymptotic time behavior.  The
agreement with the scaling law improves when the data for $n_o$ is fit
using the form (\ref{etaolaw}) with the exponents $z$, $\delta$ and
$\eta$ fixed and the coefficients $c_1$ and $c_2$ are fitted.

\begin{figure}[p]
\centerline{\psfig{file=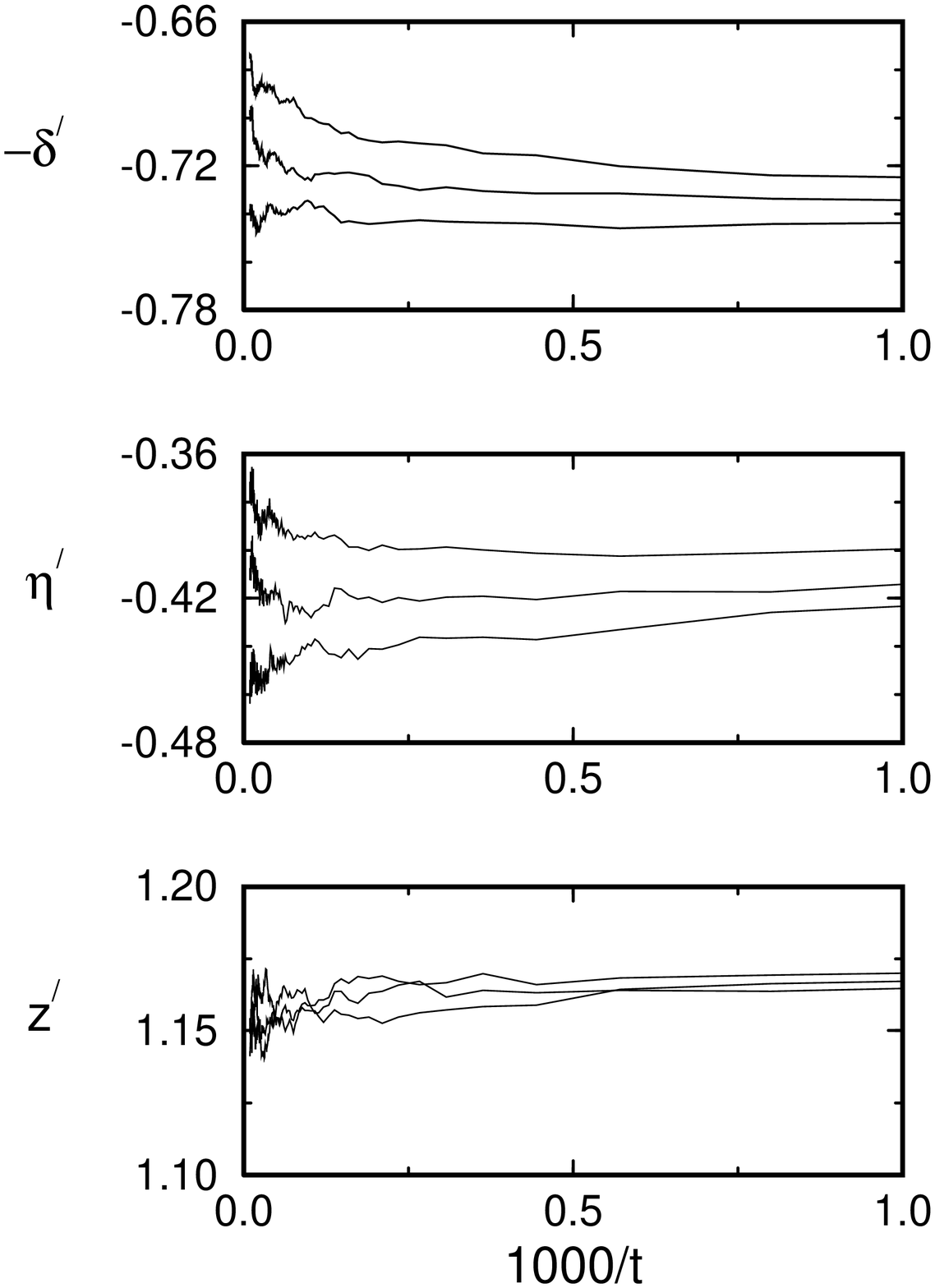,height=7.4truein}}
\caption{Effective exponents for the bicritical interface dynamics.  The
three curves correspond to three values of $r$ near the bicritical point.
From top to bottom they are $r = 0.7420,0.7435,0.7450$, with the middle
curve corresponding to the bicritical value.}
\label{interface}
\end{figure}

Figure~\ref{interface} shows the results from the interface dynamics
simulations, for three values of $r$ near $r_c = 0.7435(15)$.  From
$10^7$ independent runs of up to $10^5$ time steps we find values of
$\delta' = 0.700(10)$, $\eta' = -0.404(9)$, and $z' = 1.153(6)$.  This
type of simulation gives us information about the competition between
the growth of the two saturated phases that is not gained in the defect
dynamics type of simulations.  First, notice that $z'$, the exponent
governing the size of the interface, equals $z$.  Also, although
$\delta'$ and $\eta'$ have different values than their defect dynamics
cousins, the sum $(\delta' + \eta')$, which controls the time evolution
of the number of vacancies in just the surviving runs, is the same as
$(\delta + \eta)$  within statistical error.  This indicates that the
critical spreading of the defect is insensitive as to whether it is
bordered on each side by the same saturated phase or by two different,
though equivalent, saturated phases.  The independent dynamical
exponent $\delta'$ measured here agrees with the value found in the
three-species model\cite{shortBB,longBB}, suggesting that its value is
universal.

\section{Summary}

We have investigated the behavior of an interacting one-dimensional
adsorption-limited monomer-monomer model.  The like-neighbor repulsion
in the model leads to the presence of a large reactive region and a
bicritical point -- inaccurately termed a tricritical point in
Ref.~\onlinecite{Redner1} -- where the first-order discontinuous
transition between the two symmetric saturated phases meets the two
continuous transitions from either saturated phase to the reactive
phase.  We have used two types of epidemic simulations to study the
behavior of the system at the bicritical point.  Using the defect
dynamics simulations we determine the bicritical behavior falls in the
BAWe universality class.  With the interface dynamics we measure a new
universal number $\delta'$, related to the probability that a
minimum-width interface between the two saturated phases will avoid
collapse back to its minimum size.   It would be interesting to measure
the exponent $\delta'$ for other models in which two symmetric
adsorbing phases meet a reactive phase.  In addition, by comparing the
results from the two types of simulations we find that the time
evolution of the number of vacancies in the lattice in the surviving
runs is the same in both cases.  This suggests that the critical
spreading of a localized defect is insensitive to initial conditions,
and the characteristic fluctuations in the domain walls behave
differently than a defect in a homogeneous phase.

The failure of mean field theory to describe the qualitative features of
the phase diagram, even when clusters of triples of sites were included,
points to the importance of large domains of each species in the vicinity
of the bicritical point.  The large domains, which are clearly apparent
in static Monte Carlo simulations, are the necessary background for the
simplest dynamical variables needed to describe the bicritical point,
namely, the domain walls.  The random walks of these walls, coupled
with the fact that their number is conserved modulo 2, are the crucial
ingredients in the critical behavior at the bicritical point.

This work was supported by the National Science Foundation
under Grant No.~DMR--9408634.

\begin{table}

\begin{tabular}{cccc}
A adsorption & Rate & B adsorption & Rate \\
\hline
$VVV\to VAV$ & $p$ & $VVV\to VBV$ & $q$ \\
$VVA \to VAA$ & $p\,r$ & $VVA \to VVV$ & $q\,r$\\
$AVV \to AAV$ & $p\,r$ & $AVV \to VVV$ & $q\,r$\\
$AVA \to AAA$ & $p\,r$ & $AVA \to AVV$ & ${1\over2}q\,r$\\
              &          & $AVA \to VVA$ & ${1\over2}q\,r$\\
$VVB \to VVV$ & $p$    & $VVB \to VBB$ & $q\,r$ \\
$BVV \to VVV$ & $p$    & $BVV \to BBV$ & $q\,r$\\
$BVB \to BVV$ & ${1\over2}p\,r$ & $BVB \to BBB$ & $q\,r$\\
$BVB \to VVB$ & ${1\over2}p\,r$ &      &         \\
$AVB \to AVV$ & $p\,r$ & $AVB \to VVB$ & $q\,r$\\
$BVA \to VVA$ & $p\,r$ & $BVA \to BVV$ & $q\,r$\\
\end{tabular}

\caption{Rates for all allowed kinetic processes.  The adsorption is attempted
on the middle site in each case}
\label{ratetable}
\end{table}

\newpage

\newpage

\newpage

\newpage

\end{document}